\documentclass{scrartcl}

\usepackage[authoryear]{natbib}
\usepackage{amsmath, amssymb, amsthm}
\usepackage[ruled]{algorithm2e}

\usepackage{units}
\usepackage{booktabs}
\usepackage[toc,page]{appendix}
\usepackage{graphicx}

\DeclareMathOperator{\E}{E}
\DeclareMathOperator{\GP}{GP}

\DeclareMathOperator{\var}{var}
\DeclareMathOperator{\vectorize}{vec}

\newcommand{\N}{\mathcal{N}}
\newcommand{\newobs}[1]{\tilde{#1}}
\renewcommand{\v}[1]{\mathbf{#1}}
\newcommand{\m}[1]{\mathbf{#1}}

\title{Joint estimation of the predictive ability of experts using a multi-output Gaussian process}
\author{Oscar Oelrich$^{a,}$\thanks{Corresponding author: oscar.oelrich@stat.su.se. The computations in this paper were enabled by resources provided by the National Supercomputer Centre (NSC), funded by Linköping University, and by resources in project SNIC 2020/15-262 provided by the Swedish National Infrastructure for Computing (SNIC) at UPPMAX, partially funded by the Swedish Research Council through grant agreement no. 2018-05973.} \hspace{0.7pt} and Mattias Villani$^a$}
\date{%
    $^a$Department of Statistics, Stockholm University\\
}

\begin{document}
\maketitle

\begin{abstract}
    A multi-output Gaussian process (GP) is introduced as a model for the joint posterior distribution of the local predictive ability of set of models and/or experts, conditional on a vector of covariates, from historical predictions in the form of log predictive scores. Following a power transformation of the log scores, a GP with Gaussian noise can be used, which allows faster computation by first using Hamiltonian Monte Carlo to sample the hyperparameters of the GP from a model where the latent GP surface has been marginalized out, and then using these draws to generate draws of joint predictive ability conditional on a new vector of covariates. Linear pools based on learned joint local predictive ability are applied to predict daily bike usage in Washington DC.
\end{abstract}
\clearpage
\section{Introduction}

The out-of-sample predictive accuracy --- or \emph{predictive ability} --- of an expert has traditionally been treated as a fixed, unknown quantity \citep*{gelman_understanding_2014}. More recently, there has been an interest in modeling \emph{local} predictive ability, i.e. how the predictive ability of an expert varies locally over a space of covariates \(\v{z}\) \citep{yao_bayesian_2021, li_feature-based_2021, oelrich_local_2021, oelrich_modeling_2022}. \citet*{oelrich_local_2021} calls this covariate space the \emph{pooling space}, as a central use of local estimates of predictive ability is to pool expert predictions to create aggregate predictions \citep{bates_combination_1969,hall_combining_2007, geweke_optimal_2011, yao_using_2018, billio_time-varying_2013}.

A commonly used measure of predictive ability is the expected log predictive density (ELPD). This suggests that log predictive density evaluations, or log scores for short, can be used by an external decision maker as data to estimate local predictive ability \citep{lindley_reconciliation_1979}. However, using log scores as data is a complex task as their distribution depends on both the expert and the underlying data-generating process (DGP). For instance, when both the expert and the data-generating process are normal, the log scores will follow a scaled and translated noncentral \(\chi^2_1\) distribution \citep{sivula_uncertainty_2022}, which makes models with additive Gaussian noise a poor choice \citep{oelrich_modeling_2022}.

As discussed in \citet*{oelrich_local_2021}, specifying a model for the relationship between all the covariates and predictive ability is in itself a hard problem.
It is not uncommon to have an idea of which variables are potentially related to predictive ability, without knowing the specific structure of this relationship.
\citet*{oelrich_local_2021} therefore suggest using a  Gaussian process, as it allows us to model outputs that vary smoothly over an arbitrary set of inputs in a flexible way.

The GP suggested by \citet*{oelrich_local_2021} estimates a separate GP for each expert, hence implicitly assuming independence when forecasts are pooled based on local predictive ability.
Correctly capturing the dependence structure between experts is useful both from a pure inference standpoint, as it allows us to better understand the relationship between the predictions of the experts, as well as for creating better aggregate predictions \citep{mcalinn_multivariate_2020}.
The latter point is clearly illustrated by \citet{winkler_combining_1981}, who shows that an expert with relatively low predictive ability can nevertheless get a substantial weight in a combined forecast if the expert's forecast error is negatively correlated with the forecast errors of the other experts.

In this paper, we relax the independence assumption by implementing a multi-output GP to \emph{jointly} model the local predictive ability of a set of \(K\) experts over the pooling space.
This is done using a kernel function that is constructed by correlating a set of underlying independent GPs, one for each expert.
The construction makes it possible to use separate smoothness kernel parameters, or even completely different classes of kernels, for each expert.
It also gives us access to interpretable parameters that describe the correlation between experts.  
The multi-output model allows us to jointly infer the latent GP surfaces of all experts, which gives a more full quantification of the uncertainty of the ELPD surface of prediction pools based on the set of experts.

The joint posterior distribution of predictive ability and the kernel hyperparameters of the GP can be sampled by Hamiltonian Monte Carlo (HMC).
However, as the dimension of the sampled joint distribution grows linearly with the sample size, estimation time quickly becomes a problem.
To get around this, \citet*{oelrich_local_2021} apply a power transformation to the log scores to obtain approximate normality with homoscedastic variance.
Following this transformation, a much faster posterior sampler that marginalizes out the latent GP surface can be used to generate draws from the hyperparameters of the kernel. The hyperparameter draws can in turn can be used to generate draws from the joint posterior of predictive ability conditional on a new vector of values \(\mathbf{z}\) using standard methods \citep[Chapter~2]{rasmussen_gaussian_2006}.
We show that the same approach can be used for the multivariate model proposed here.

The paper proceeds as follows.
Section \ref{sec_joint_pred_abil} extends the framework of local predictive ability of an expert from \citet{oelrich_modeling_2022} to the joint local predict ability of a set of \(K\) experts, and introduces a multi-output GP on power-transformed log scores as a method to model joint local predictive ability.
Section \ref{sec_simstud} illustrates the multi-output GP and compares it with the single output approach with a simple simulation study.
Section \ref{sec_empirical} applies the multi-output GP to a bike sharing dataset and uses the learned predictive ability for pooling predictions. Section \ref{sec_conclusions} concludes.

\section{Joint predictive ability}\label{sec_joint_pred_abil}

This section defines local predictive ability within a decision maker framework \citep*{oelrich_local_2021}, discusses the implications of using log scores as data to estimate local predictive ability, and proposes a multi-output Gaussian process as a model for estimating joint local predictive ability.

\subsection{Local predictive ability}

Consider a decision maker \((\mathcal{DM})\) who has access to a set of \(K\) experts, either in the form of models or skilled humans, and wants to evaluate the predictive accuracy of these experts, based on their past predictions \citep*{lindley_reconciliation_1979}. These past predictions are available to the decision maker in the form of log predictive density scores, $\ell_{ik} = \log p_k(y_i \mid \v{y}_k)$, where \(p_k(y_i \mid \v{y}_k)\) is the predictive density of expert \(k\) for observation \(y_i\), conditional on the data \(\v{y}_k\) used by expert $k$. This situation occurs whenever a set of formal predictive statistical models are compared, but also increasingly with human experts that provide  probabilistic assessments, for example in economics \citep{croushore2019fifty,de2019twenty} and meteorology \citep{council2008enhancing}.

The decision maker defines the predictive ability of expert \(k\) as the expected log predictive density of that expert for a new data point \(y\) from the data-generating process \(p_*(y)\), \(\E_{p_*(y)}\left[\log p_k(y \mid \v{y}_k)\right]\) \citep{gelman_understanding_2014}. We assume that the decision maker has access to multiple experts predicting a univariate observations, but the extension to multivariate observations is immediate.
The decision maker further believes that the predictive ability of the experts vary over a set of pooling variables \(\v{z}\), and define the \emph{local predictive ability} \citep{oelrich_local_2021} of expert \(k\) at \(\v{z}\) as
\begin{equation}\label{lelpd}
    \eta_k(\v{z}) \equiv
    \operatorname{E}_{p_*(y)}
    \left[
        \log p(y\mid \v{y}_k, \v{z})
    \right]
     =
    \int_{-\infty}^{\infty} 
    \left[
        \log p(y\mid \v{y}_k, \v{z})
    \right]
     p_*(y)\,dy.
\end{equation}

\subsection{Models for univariate log scores}

In the approach described above, the log scores of the experts are used as data to model \(\eta_k(\v{z})\).
The appropriate model for the log scores depends on the predictive distribution of the expert as well as the data-generating process (DGP).

We focus here on the ubiquitous scenario where the experts' predictive distributions and the data generating process are all Gaussian, i.e. that the predictive distribution of expert \(k\) for observation \(y_i\) is \(\N(\mu_{ik}, \sigma^2_{ik})\) and that the DGP for the same observation is given by \(\N(\mu_{i*}, \sigma^2_{i*})\).
It then follows that the log scores can be rewritten as scaled and translated noncentral \(\chi^2_1\) variables \citep{sivula_uncertainty_2022, oelrich_modeling_2022}
\begin{equation}\label{eq:nc_lincomb}
    \ell_{ik} \overset{d}{=} a_{ik} - b_{ik}\cdot X_{ik},
\end{equation}
where ${X_{ik} \sim \chi^2_1(\lambda_{ik})}$ with non-centrality parameter ${\lambda_{ik}=(\mu_{i*}-\mu_{ik})^2/\sigma_{i*}^2}$, ${b_{ik}=\sigma_{i*}^2/(2\sigma_{ik}^2)}$ and ${a_{ik}=-\frac{1}{2}\log(2\pi\sigma_{ik}^2)}$.

Note that $a_{ik} = -\frac{1}{2}\log(2\pi\sigma_{ik}^2)$ is determined by the expert's predictive distribution which we assume to be given, so we can define $\ell'_{ik} \equiv -(\ell_{ik}-a_{ik})$ and reformulate the model in \eqref{eq:nc_lincomb} as
\begin{equation}\label{eq:modelwithdifferent_bik}
\ell'_{ik} \mid \lambda_{ik},b_{ik} \sim \chi^2_1(\lambda_{ik}, b_{ik}),
\end{equation}
where $\chi^2_1(\lambda, b)$ is used to denote a \emph{scaled} non-central $\chi^2$ with one degree of freedom, non-centrality parameter $\lambda$ and scale parameter $b$. Following  \citet{oelrich_modeling_2022}, we assume that $b_{ik} = b_k$ for all $i = 1,\ldots,n$, i.e. that the variance ratio $\sigma_{i*}^2/\sigma_{ik}^2$ is constant over observations for a given expert. 

Our end goal is to model predictive ability as it varies over the pooling space.
Since the local predictive ability of an expert is, by definition, the conditional expectation of the log scores, we can model \(\eta(\v{z})\) indirectly by first modeling the log scores.
We can then extract the posterior distribution of the mean from the model of the log scores.
Equation \eqref{eq:modelwithdifferent_bik} suggests that a suitable candidate would be to model the transformed log scores using a scaled non-central \(\chi^2_1\) distribution.

The noncentral \(\chi^2_1\) model can be made local by letting the non-centrality parameter \(\lambda_{k}\) depend on the pooling variables.
While the specific nature of how \(\lambda_{k}\) varies over the pooling space is ultimately up to the decision maker, a flexible model that makes minimal assumptions about the structural form is suitable when the decision maker has a weak prior understanding of the specific way in which predictive ability varies over the pooling space.
For example, in \citet{oelrich_modeling_2022}, transformed log scores are modeled as \(\ell'_{ik} \sim\chi^2_1(\lambda_{k}(\v{z}_i), b_k)\) where \(\log \lambda_k(\v{z})\) is given by a Gaussian process.

A Gaussian process generalizes the Gaussian distribution from a random variable to a stochastic process, resulting in a probability distribution over functions \citep*{rasmussen_gaussian_2006}.
A Gaussian process is a collection of random function values with different inputs \(\{f(\v{z}_1), f(\v{z}_2), \dots\}\), where each finite subset follows a multivariate Gaussian distribution.
The covariance matrix of each subset is defined by a covariance function, which in turn is a function of the inputs.
The specific form of this covariance function will determine our prior regarding how the outputs vary over the input space.
For example, the covariance function we use in the examples and applications in this paper, the squared exponential with automatic relevance determination \citep{neal_bayesian_1996}, is given by
\begin{equation}\label{squared_exponential}
    g_{\operatorname{SE}}(\v{z}_i, \v{z}_j) =
    \alpha^2 \exp \left(
        -\frac{1}{2} 
        (\v{z}_i - \v{z}_j)
        (\operatorname{diag}(\boldsymbol{l}^{-2}))
        (\v{z}_i - \v{z}_j)
    \right),
\end{equation}
where \(\boldsymbol{l}=(l_1, \dots, l_P)\) is a vector of \emph{length scales} corresponding to the \(P\) inputs, and \(\alpha^2\) denotes the signal variance.
The automatic relevance determination (ARD) part of the covariance function refers to that the length scales are allowed to be different for different input variables.
For example, if the input \(\v{z}\) is two-dimensional, and the first length scale is much smaller than the second, this would mean that the covariance function is much more sensitive to changes in the first input.
An input with a large length scale will have virtually no effect on the covariance function, which can be interpreted as the input lacking relevance for the response \citep{rasmussen_gaussian_2006}.

The model \(\ell'_{ik} \sim\chi^2_1(\lambda_{k}(\v{z}_i), b_k)\), where \(\log \lambda_k(\v{z})\) is given by a Gaussian process in \citet{oelrich_modeling_2022}, lacks a closed form posterior and the posterior thus needs to be sampled using a Monte Carlo method such as Hamiltonian Monte Carlo (HMC).
A drawback of this model is that it requires sampling the high-dimension GP surface and the kernel hyperparameters jointly, which is a parameter space that can be of high dimension. So while HMC sampling for the \(\chi^2_1\) GP-based model is efficient, it is still slow in even moderately sized datasets since \(\boldsymbol{\lambda}_{k} = (\lambda_{k}(\v{z}_1),\ldots,\lambda_{k}(\v{z}_n))^\top \) grows linearly with the sample size \citep{oelrich_modeling_2022}.
As a solution, \citet{oelrich_modeling_2022} suggest applying a cube root transformation to the transformed log scores to obtain approximate normality with constant variance. 
Following this second transformation, a Gaussian process regression with Gaussian noise can be used, a model which has the advantage that it can be sped up significantly by analytically integrating out the latent GP surface.
We will refer to this model, a GP based on cube-root transformed data, as the \(\GP(\nicefrac{1}{3})\) model.

\citet{oelrich_modeling_2022} use the \(\GP(\nicefrac{1}{3})\) model to estimate the local predictive ability individually for a set of experts.
When these local predictive abilities are used for pooling, two independence assumptions are implied: independence in predictive ability, and independence in the errors conditional on predictive ability.
In the following section, we propose a generalization of the \(\GP(\nicefrac{1}{3})\) model to a multi-output GP that relaxes both of these assumptions.

\subsection{The multi-output \(\operatorname{GP}(\nicefrac{1}{3})\) model}

Let \(\ell_{ik}^{\prime\prime} = \left(a_{ik} - \ell_{ik}\right)^{\nicefrac{1}{3}}\) denote the cube-root transformed log scores. The distribution of \(\ell_{ik}^{\prime\prime}\) for \(K\) experts jointly, \(\boldsymbol{\ell}_{i}^{\prime\prime} = (\ell_{i1}^{\prime\prime},\ldots,\ell_{iK}^{\prime\prime})^\top\), can then be modeled using a multi-output GP with Gaussian noise
\begin{align}
\begin{split}
        \boldsymbol{\ell}''_{i} \mid 
        \v{f}(\v{z}_i), 
        \boldsymbol{\Sigma}
       &\sim
        \N\left(\v{f}(\mathbf{z}_i), \boldsymbol{\Sigma}\right)\\
        \mathbf{f}(\v{z}) \mid \boldsymbol{\Theta}, \m{C}
        &\sim
        \operatorname{GP}\left(\boldsymbol{\mu}(\v{z}), g(\v{z}, \v{z}^\prime)\right),
\end{split}
\label{multioutGP}
\end{align}
where \(\v{f}(\v z) = (f_1(\v z),\ldots,f_K(\v z))^\top\) is a multi-output GP with hyperparameters \(\boldsymbol{\Theta} = \{\boldsymbol{\theta}_1\),\ldots,\(\boldsymbol{\theta}_K\}\) including a vector of parameters \(\boldsymbol{\theta}_k\) for the mean and covariance function of each expert as well, and \(\m{C}\) is a \(K \times K\) matrix that will be used below to correlate the experts' abilities \(\mathbf{f}\). The noise covariance matrix \(\boldsymbol{\Sigma}\) is here taken to be a general positive definite matrix, but may also be restricted to a diagonal matrix.
The goal of the model in \eqref{multioutGP} is to capture correlation of the log scores between experts.
This correlation is coming from two sources: in the noise distribution via \(\boldsymbol{\Sigma}\), and in the underlying GP prior for \(\v{f}(\v{z})\) with correlation generated by \(\m{C}\), which we now describe.

The model in \eqref{multioutGP} for all log scores in the training sample can be written
\begin{align}\label{multimodel}
    \underset{n\times K}{\boldsymbol{\ell}''} &=
        \underset{n\times K}{\m{F}} +
        \underset{n\times K}{\m{E}},
\end{align}
where \(\boldsymbol{\ell}''\) is the matrix consisting of the transformed log scores of \(K\) experts at \(n\) time points, \(\m{F}\) is determined by the multi-output GP, and \(\m{E}\) is a matrix of Gaussian noise terms. 

Reformulating the model in vectorized form, we can write the vector of errors as
\({\vectorize \m E = \boldsymbol{\Sigma} \otimes \m{I}_n}\). By letting \(\boldsymbol{\Sigma}\) be an arbitrary covariance matrix we allow correlation in the error terms across experts, but assume independence between observations.

Modeling the correlation in the underlying predictive ability is done via the GP prior.
A first attempt for a model for \(\m{F}\) is a matrix normal 
\begin{equation}\label{matrixnormal}
     \m{F} \sim \operatorname{\mathcal{MN}}(\v M, \v G,\v \Omega),
\end{equation}
i.e. that \(\vectorize \m{F}\) follows a multivariate normal with Kronecker-structured covariance matrix \(\m\Omega \otimes \m G\).
The matrix $\m G$ models the correlation between rows of $\m F$, which correspond to observations, and is therefore typically a covariance matrix generated from a kernel function, for example the squared exponential in \eqref{squared_exponential}.
The matrix $\m \Omega$ models the correlation between columns, in our case experts, and can most naturally be taken as a general positive definite covariance matrix. The signal variances in the squared exponential kernel, $\alpha$, would then be set to $1$ for each expert.

While the matrix normal model in \eqref{matrixnormal} has some intuitive appeal and is easy to work with from a computation perspective, it assumes that the columns of $\m F$ all follow the same multivariate normal distribution, which puts the restriction on our model that all experts have identical covariance function, i.e. the same covariance matrix $\m G$.
To be able to specify separate covariance functions for each expert we instead induce a prior for \(\m F\) by correlating a set of \(K\) underlying independent GPs, $\v h_1,\ldots,\v h_K$, each with its own kernel function \citep{teh_semiparametric_2005}
\begin{equation}\label{tilde_ver}
    \m F  = \m H \m C,
    \qquad
    \m H = \left[ \v h_1 \dots \v h_K \right],
    \qquad
    \v h_k \overset{\mathrm{indep}}{\sim} \N\left[\v 0, \operatorname G_k(\m Z, \m Z)\right],
\end{equation}
where \(\operatorname G_k(\m Z, \m Z^{\prime})\) is the covariance matrix for expert \(k\) generated from kernel function $g_k(\v z,\v z^\prime)$ and $\m C$ is a $K \times K$ matrix that generates the dependence between the $\v f_k$ for the experts.
We set the signal standard deviation to unity in all kernels, i.e. $\alpha_1=\cdots=\alpha_K=1$ in the squared exponential kernel in \eqref{squared_exponential} so that the scales of $\m F$ is generated by the $\m C$ matrix.
While the decision maker does not need to use the same pooling variables for each expert, we let the matrix of pooling variables \(\m Z\) contain the pooling variables for all expert to simplify notation.

The model in \eqref{tilde_ver} allows the decision maker to specify completely different covariance functions for the experts.
Importantly, this allows the length scales to differ between the expert in the case where an ARD kernel is used, but it is also possible to use, for example, a Matérn kernel \citep{rasmussen_gaussian_2006} for some of the experts and squared exponential for others.
It is also straightforward to force a subset of experts to share the exact same covariance function.

The covariance function for \(\m F\) in \eqref{tilde_ver} is a weighted sum of the covariance functions of the \(K\) experts
\begin{equation}\label{MV_covform}
    \operatorname{cov}(f_{k}(\v z_i), f_{l}(\v z_j)) = \sum_{s = 1}^K c_{sk}c_{sl} \mathrm{g}_s(\v z_i, \v z_j),
\end{equation}
where \(c_{kl}\) is element \((k,l)\) of the matrix \(\m C\).
To fully specify the multi-output \(\GP(\nicefrac{1}{3})\) model --- which we will denote by \(\operatorname{multi-GP}(\nicefrac{1}{3})\) --- we need to specify the covariance function of each expert, including priors for any unknown hyperparameters, as well as a prior for the matrix \(\m C\) that describes the between-expert covariances.

Note that while we choose to select a number of underlying processes equal to the number of experts, there is nothing that prevents us from creating our multi-output process by correlating an arbitrary number of latent GPs.
A common justification for using several models is that it gives a more holistic quantification of uncertainty by capturing the structural uncertainty of the parametric form of the models, rather than just the parametric uncertainty within models \citep{draper_assessment_1995}. 
From this perspective, including several models that share the majority of structural DNA can be problematic, for example in Bayesian model averaging \citep{hoeting_bayesian_1999}, where adding an effective "copy" of an already existing model while using a uniform prior will essentially double the posterior model weight of that model.
When modeling the joint predictive ability of a large pool of highly correlated experts it is possible that many of the elements of \(\m C\) will be close to zero, in which case restricting the number of underlying GPs could simplify the model and speed up calculations.

\subsection{Inference for the multi-output \(\operatorname{GP}(\nicefrac{1}{3})\) model}

When modeling local predictive ability, the end goal is typically the posterior of the ELPD for all experts at a new point \(\newobs{\mathbf{z}}\) in the pooling space, $\boldsymbol{\eta}(\v{z}) = (\eta_1(\newobs{\v{z}}),\ldots,\eta_K(\newobs{\v{z}}))^\top$, i.e. the mean of \(\boldsymbol{\ell}\) at \(\newobs{\mathbf{z}}\). Since \(\E(\boldsymbol{\ell}_{i})=\left(\E(\ell_{i1}), \dots, \E(\ell_{iK}) \right)^{\intercal}\) and \(\ell_{ik} = a_{ik} - \left(\ell_{ik}^{\prime\prime}\right)^3\), we can use a formula for the third moments of Gaussians \citep{oelrich_modeling_2022} to obtain the local predictive ability from model \eqref{multioutGP} pointwise as
\begin{equation}\label{eq:ELPDcuberoot}
    \eta_k(\newobs{\v{z}}) = \newobs{a}_k - f_k^3(\newobs{\v{z}}) - 3f_k(\newobs{\v{z}})\boldsymbol{\Sigma}_{k, k},
\end{equation}
where \(\boldsymbol{\Sigma}_{k, k}\) denotes the \(k\)th diagonal element of the noise covariance matrix, and \(f_k(\newobs{\v{z}})\) the \(k\)th output of the GP at \(\newobs{\v{z}}\).

According to Equation \eqref{eq:ELPDcuberoot} we obtain the posterior for $\boldsymbol{\eta}(\newobs{\v{z}})$ through the joint posterior ${p\left(\v{f}(\newobs{\v{z}}),\boldsymbol{\Sigma}, \boldsymbol{\Theta}, \m C \mid \boldsymbol{\ell}^{\prime\prime}\right)}$, where \(\boldsymbol{\ell}^{\prime\prime}\) is the $n \times K$ matrix of observed (transformed) log scores for all experts.
This posterior is expensive to sample from, however.
The cube root transformation makes it possible to extend the results in \citet{oelrich_modeling_2022} to analytically integrate out $\m F$ in the training data , and obtain the marginal posterior of $\boldsymbol{\Sigma}$ and the GP hyperparameters ${\boldsymbol{\Theta} \text{ and } \m{C}}$ in closed form using a multivariate version of Eq.~2.8 in \citet{rasmussen_gaussian_2006}
\begin{equation}\label{eq:marglikegaussianGPreg}
    p\big(\v \Sigma,\boldsymbol{\Theta}, \m C \mid \boldsymbol{\ell}^{\prime\prime}\big) = \int p\big(\m F,\boldsymbol{\Sigma},\boldsymbol{\Theta}, \m C \mid \boldsymbol{\ell}^{\prime\prime}\big)\,d\m F.
\end{equation} 
This lower-dimensional marginal posterior can be sampled with Hamiltonian Monte Carlo (HMC) using Stan \citep{stan_development_team_stan_2022, gabry2020cmdstanr} in a fraction of the time that it takes to sample the full joint posterior.

For each HMC sampled \((\boldsymbol{\Theta}, \m C, \boldsymbol{\Sigma})\) from \eqref{eq:marglikegaussianGPreg} we can now easily sample from the distribution of \(\v{f}(\newobs{\v{z}})\) using \citep*{rasmussen_gaussian_2006}
\begin{equation}\label{eq:predictivefstar}
    \v{f}(\newobs{\v{z}})
    \mid 
    \boldsymbol{\ell}^{\prime\prime}, \m{Z}, \newobs{\v{z}}, \boldsymbol{\Sigma}, \boldsymbol{\Theta} , \m C
    \sim 
    \N\left[
        \bar{\tilde{\v{f}}}, \operatorname{var}(\tilde{\v{f}}) 
    \right],
\end{equation}
where \(\m{Z}\) is the matrix of pooling variables in the training data, and
\begin{align*}
    \bar{\tilde{\v{f}}} &= \boldsymbol{\mu}(\newobs{\v z}) + 
    \operatorname{G}(\newobs{\v{z}}, \m{Z})
    \left[
        \operatorname{G}(\m{Z}, \m{Z}) + \boldsymbol{\Sigma} \otimes \m I_n
    \right]^{-1} (\vectorize\boldsymbol{\ell}^{\prime\prime} - \boldsymbol{\mu}),\\
    \var(\tilde{\v{f}}) &=
    \operatorname{G}(\v{z}, \newobs{\v{z}}) -
    \operatorname{G}(\newobs{\v{z}}, \m{Z})
    \left[
        \operatorname{G}(\m{Z}, \m{Z}) + \boldsymbol{\Sigma} \otimes \m I_n
    \right]^{-1}
    \operatorname{G}(\m{Z}, \newobs{\v{z}}),
\end{align*}
where, for example, \(\operatorname G(\m{Z}, \m{Z})\) is the \(Kn\times Kn\) matrix with covariances of the form \(\operatorname{cov}(f_k(\v{z}_i),f_l(\v{z}_j))\) for experts \(k\) and \(l\) and observations \(i\) and \(j\) in the training data.
The posterior draws of \(\boldsymbol{\Sigma}\) from \eqref{eq:marglikegaussianGPreg} and \(\v{f}(\newobs{\v{z}})\) from \eqref{eq:predictivefstar} can finally be inserted into \eqref{eq:ELPDcuberoot} to obtain draws from the posterior of the local ELPD \(\boldsymbol{\eta}(\newobs{\mathbf{z}})\).
The method is summarized in Algorithm \ref{alg:GPcuberootjoint}. 

\begin{algorithm}
    \caption{Sampling from the joint posterior of local predictive ability of the Multi-\(\GP\left(\nicefrac{1}{3}\right)\) model}\label{alg:GPcuberootjoint}
    \KwIn{Number of posterior draws $M$.
    }
    \vspace{2mm}
  
  Generate \(M\) draws \(\{\boldsymbol{\Sigma}^{(j)}, \boldsymbol{\Theta}^{(j)}, \m{C}^{(j)}\}_{j=1}^M\) from \(p\left(\boldsymbol{\Sigma},\boldsymbol{\Theta}, \m{C} \mid \boldsymbol{\ell}^{\prime\prime}\right)\) in \eqref{eq:marglikegaussianGPreg} using HMC.
  \vspace{1mm} \\
  \For {\(j = 1,\dots, M\)}{
  Sample the posterior $\v{f}^{(j)}(\newobs{\v{z}}) \mid 
    \boldsymbol{\ell}^{\prime\prime}, \m{Z}, \newobs{\v{z}}, \boldsymbol{\Sigma}^{(j)}, \boldsymbol{\Theta}^{(j)}, \m{C}^{(j)}$ in \eqref{eq:predictivefstar}\\
    Compute joint local predictive ability $\boldsymbol{\eta}^{(j)}(\newobs{\v{z}})$ from \eqref{eq:ELPDcuberoot}
  \\ \vspace{1mm}
  }
  \vspace{1mm}
 \KwOut{Sample from the joint local predictive ability posterior: $\boldsymbol{\eta}^{(1)}(\newobs{\v{z}}), \boldsymbol{\eta}^{(2)}(\newobs{\v{z}}), \ldots, \boldsymbol{\eta}^{(M)}(\newobs{\v{z}})$.}
\end{algorithm}

\section{A simulation-based exploration of the multi-output model}\label{sec_simstud}

The aim of this section is to examine some of the properties of the multi-output \(\operatorname{GP}(\nicefrac{1}{3})\) model based on simulated data, and to compare it with the single output version.

The questions we focus on are: i) can the \(\operatorname{multi-GP}(\nicefrac{1}{3})\) model leverage the correlation between the predictive ability of the experts to make better estimates of local predictive ability, ii) does the multi-output model still perform acceptably as compared with the single output models when there is zero correlation between the experts, iii) does the multi-output model manage to correctly identify the relevant pooling variables for each expert, as measured by the marginal posterior distribution of the length scales?

The answers will clearly depend on the specific simulation setup. For example, the stronger the correlation between the predictive ability of the experts, the easier it is for the multi-output model to outperform the single output ones. Further, a more complicated model will typically struggle under smaller sample sizes, so we explore the questions posed above at sample sizes of \(100\) and \(400\) observations per expert. 

\subsection{Simulation setup}

The primary purpose of this simulation study is to explore the ability of the multi-output \(\GP(\nicefrac{1}{3})\) model to estimate predictive ability at a new point in the pooling space.
In actual applications this will be done using data in the form of transformed log scores, which will, under the assumption of normal DGP and expert, follow a scaled \(\chi^2_1(\lambda)\) distribution.
\begin{equation}
    \ell^{\prime} \overset{d}{=} b \chi^2_1(\lambda).
\end{equation}

We will generate pseudo log-scores as
\begin{equation}\label{pseudolog}
    \ell^{\prime}_{ik} = b_k (y_i-z_{ik})^2, \hspace{0.5cm} \text{ for } i=1,\ldots,n,
\end{equation}
where \(b_k=\nicefrac{1}{3}\) is kept fixed and \(y_i\sim\N(0, 1)\).
In practice, this means that we induce variation in the predictive ability of each expert over the pooling space by letting the difference between the predictive mean of each expert and that of the DGP vary in a systematic way over the pooling space.
We let this difference in mean depend on a different pooling variable for each expert.
Specifically, the conditional mean of the log scores for the \(k\)th expert depend only on the \(k\)th pooling variable \(z_k\) in \eqref{pseudolog}.
Note that, conditional on \(z_{ik}\), the expectation of \eqref{pseudolog} is \( \eta_k(\v z_k)=b(1+z_{ik}^2)\).
To keep things simple we only use \(K=2\) experts. 

The fitted multi-output model deviates from the DGP in two ways.
First, the dependence between the experts in the DGP is a one-dimensional dependence driven by a single random output \(y\) which is used to compute the log scores for all experts; hence the fitted model is overparameterized with a much richer source of dependence coming from both the signal and the noise.
Second, the cube root transformation in the fitted model will only approximately give a homoscedastic Gaussian model \citep{oelrich_modeling_2022}.

While the predictive ability of each expert only depends on one of the pooling variables, this information is typically unknown, so we model local predictive ability over \(\mathbf{z} = (z_1, z_2)\) for both experts.
The pooling variables are generated using a multivariate normal distribution with standard normal marginals.
To generate correlated expert predictions, a covariance of \(0.7\) between \(z_1\) and \(z_2\) is used.

\begin{figure}
    \centering
    \includegraphics[width=0.65\textwidth]{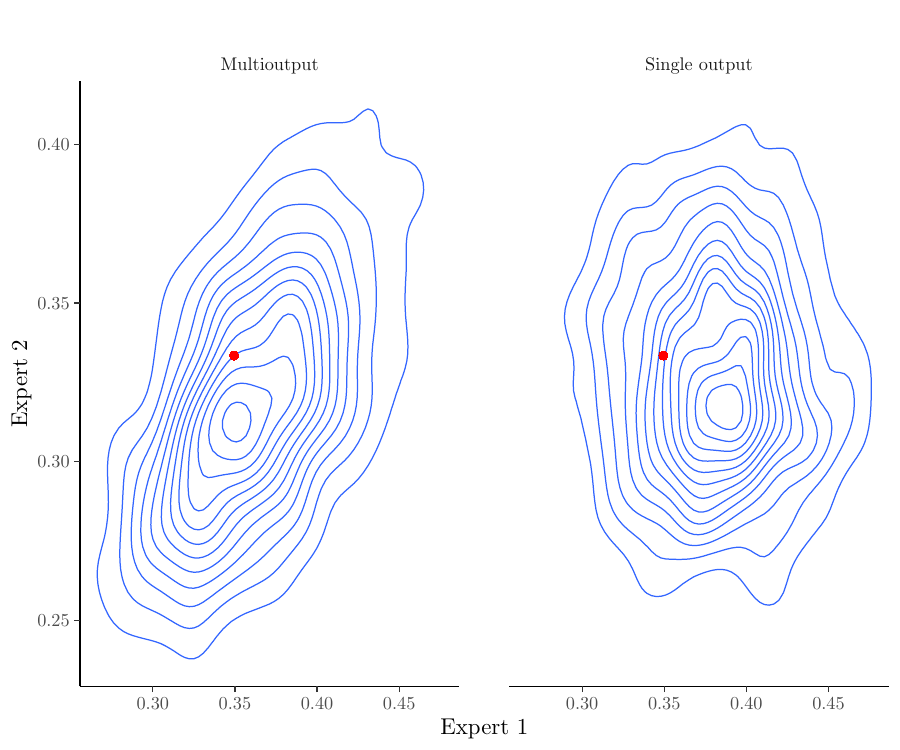}
    \caption{Posterior distribution of joint local predictive ability for a correlated data-generating process. The red dot indicates the true value.}
    \label{fig:ss1_corr}
\end{figure}

The primary quantity of interest is the joint posterior distribution of predictive ability at a new point in the pooling space \(\newobs{\v{z}}\), i.e. \(\boldsymbol{\eta}(\newobs{\v{z}})=(\eta_1(\newobs{\v{z}}),\eta_2(\newobs{\v{z}}))\).
Figure \ref{fig:ss1_corr} illustrates this distribution for one of the datasets in the simulation study, together with the true value, for both models when \(z_1\) and \(z_2\) are correlated. 
To quantify the difference in performance between the multi-output \(\operatorname{GP}(\nicefrac{1}{3})\) model and the two single output models, we generate \(100\) datasets each for correlated and uncorrelated experts.
For each generated dataset, we calculate the joint posterior predictive density of the true predictive ability at a randomly  selected point \(\newobs{\v{z}}\).
The results are summarized in Table \ref{table:lsls_small}.
As expected, the \(\operatorname{multi-GP}(\nicefrac{1}{3})\) model is slightly outperformed by the single output models when the data-generating process is uncorrelated.
When the data-generating process is correlated, the \(\operatorname{multi-GP}(\nicefrac{1}{3})\) outperforms the single output models.
\begin{table}
    \centering
    \begin{tabular}{lcccc}
    \toprule 
     & \multicolumn{2}{c}{n = 100}& \multicolumn{2}{c}{n = 400} \\
     \cmidrule(r){2-3} \cmidrule(r){4-5}
            & Correlated & Uncorrelated & Correlated & Uncorrelated \\
    \midrule
     Multi-output    & \(\boldsymbol{1.33}\)   & 1.11 & \(\boldsymbol{2.53}\) & 2.43 \\
     Single output  & 1.17   & \(\boldsymbol{1.14}\) & 2.34 & \(\boldsymbol{2.53}\) \\ 
    \bottomrule
    \end{tabular}
    \caption{Mean log predictive score for joint estimates of predictive ability for \(100\) datasets. Best model for each setup are indicated by bold fonts.}
    \label{table:lsls_small}
\end{table}

\begin{figure}
    \centering
    \includegraphics[width=1\textwidth]{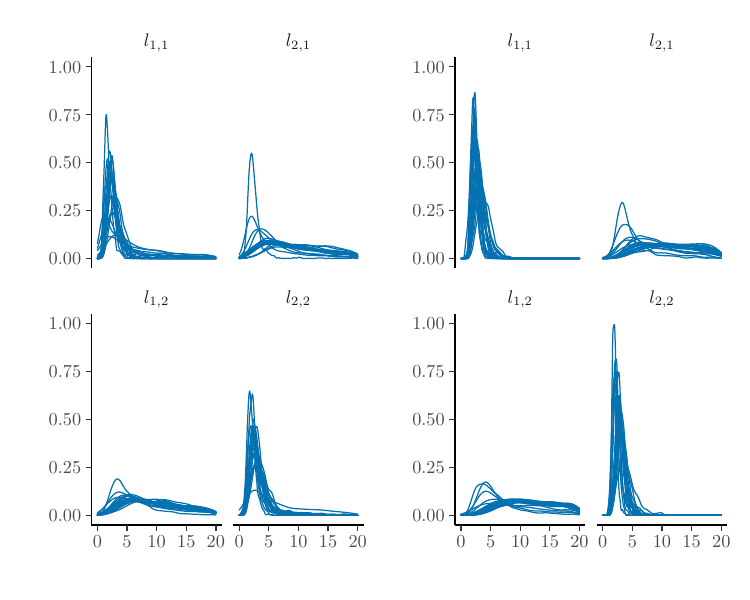}
    \caption{Marginal posterior distributions of the length scales in the multi-GP($\nicefrac{1}{3}$) model in $20$ randomly selected datasets of sample size $n=100$ (left) and $n=400$ (right).}
    \label{fig:ss2_lsls_101}
\end{figure}

As a secondary but related concern, we wish to explore how well the multi-output \(\GP(\nicefrac{1}{3})\) model manages to identify the relevant pooling variables for each expert.
The way the simulation is set up, the model is successful in identifying relevance whenever the posterior mean of the first length scale of Expert 1, \(l_{1,1}\), is concentrated on small values while its second length scale \(l_{2,1}\) is centered on larger values (ability of Expert 1 depends on $z_1$, but not on $z_2$), and vice versa for Expert 2.
The posterior distribution of the length scales from \(20\) of the generated datasets are shown in Figure \ref{fig:ss2_lsls_101}, with the left side corresponding to datasets of size \(n=100\) and the right to datasets of size \(n=400\). The figure does indeed show that the posteriors for \(l_{1,1}\) and \(l_{2,2}\) are generally more concentrated on smaller values than the posteriors for \(l_{2,1}\) and \(l_{1,2}\).
The differences in posteriors between relevant and irrelevant inputs clearly increase with the size of the dataset, but the model works well even for smaller datasets.

\section{Predicting bike-sharing utilization rates in Washington D.C.}\label{sec_empirical}

The last decade has seen an explosion in companies offering short-term rentals of final mile transportation, such as bikes or electric scooters.
Key for the success of such a business is to be able to correctly predict demand.
In this section we use the multi-output GP model to estimate the joint one-step-ahead local predictive ability of three experts predicting bike rentals based on the data in \citet*{fanaee-t_event_2014}.
We then use the estimated joint local predictive ability from the multi-output GP to form aggregate predictions, and compare the out-of-sample performance of these predictions with predictions made using multiple single-output GPs, as well as a selection of other methods from \citet{oelrich_modeling_2022}.

\subsection{Inferring local predictive ability}

The dataset in \citet*{fanaee-t_event_2014} includes the daily number of bike rentals over the two year period from January 1, 2011, to December 31, 2012.
The dataset further contains several covariates relating to the weather, as well as indicators for official US holidays and an indicator for if the day in question is a workday.

Following \citet{oelrich_local_2021}, we use three experts to generate predictions: a Bayesian regression model (BREG), a Bayesian additive regression tree model (BART), and a Bayesian linear regression model with stochastic volatility (SVBVAR).
These experts use several weather-related variables, an indicator for season, the number of rentals the previous day, and indicators for workday and holiday as covariates.
For further discussion of the data see \citet*{oelrich_local_2021}.

As pooling variables the decision maker uses humidity, wind speed, and temperature from the dataset in \citet*{fanaee-t_event_2014}.
The decision maker also decides to use a variable she calls \emph{family holiday} which is not included in the dataset, or in the estimation of any of the experts, which takes the value \(1\) at Thanksgiving and Christmas (Eve and Day) \citep*{oelrich_local_2021}.
This variable is included to represent the decision makers belief that there are certain stay-at-home holidays where the demand for rental bikes is almost non-existent.
As neither of the experts have access to this variable, she expects their predictive ability to vary significantly with regards to this variable. 

To estimate the joint predictive ability of all three experts we use the \(\operatorname{multi-GP}(\nicefrac{1}{3})\) model, with a squared exponential automatic relevance determination (ARD) kernel \citep*{rasmussen_gaussian_2006}.
To allow for the possibility that some pooling variables are unimportant to some experts we use a fairly flat prior for the length scales, \(\operatorname{Cauchy}(0, 5)\), restricted to the interval \((0, 100)\).
For the \(\m C\) matrix we use a \(\mathrm{LKJ}(\eta = 3)\) prior \citep{lewandowski2009generating}, and zero-truncated \(\N^+(0, 1)\) priors for the signal and error term variances.

\begin{figure}
    \centering
    \includegraphics {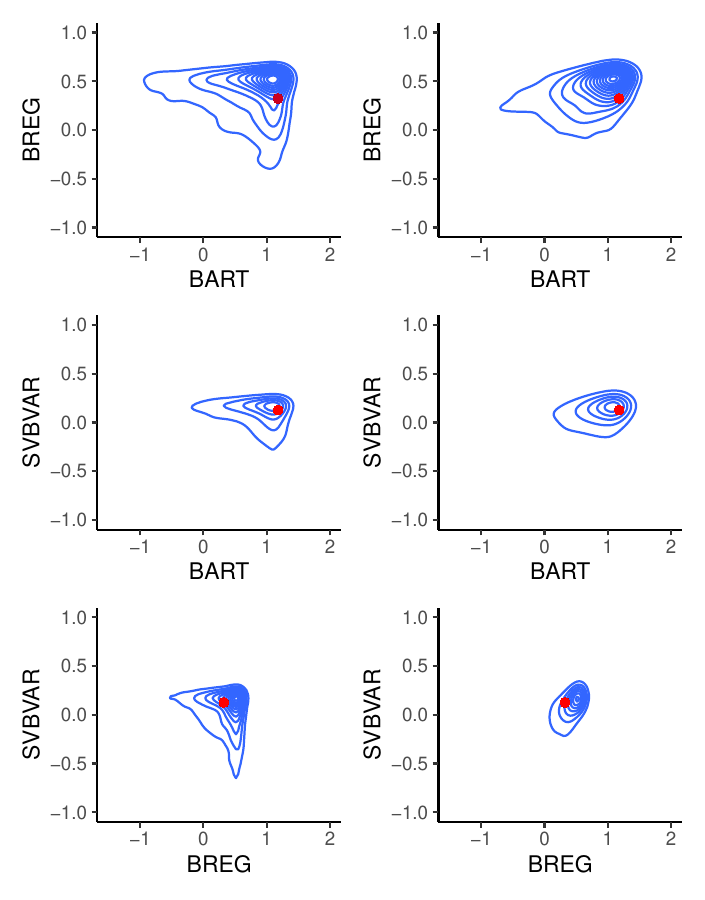}
    \caption{Joint predictive distributions of log scores for the last data point in the bike-sharing data. The graphs show kernel density estimates for univariate (left side) and multivariate (right side) Gaussian processes. The observed log scores are marked with a red dot.}
    \label{fig:ls_preddist}
\end{figure}

Figure \ref{fig:ls_preddist} displays the bivariate joint predictive distribution of each expert pair at the last point in the data set, with the first column showing the joint predictive distributions from single output GPs --- implicitly assuming independence --- and the second column showing the joint distributions based on the multi-output model.
The multi-output model captures positive correlations between all experts, especially between the Bayesian regression model and the stochastic volatility model.
The smaller correlations with BART is is not surprising, since the BART model is a quite different model with possibly highly nonlinear mean function.
Note also that the predictive variances for the multi-output distributions are \(10\)--\(20\%\) lower at this time point.

\subsection{Using joint local predictive ability for forecast combination}

There are many reasons to model joint predictive ability.
For example, it can be done purely for inference purposes, such as when we are interested in examining how the quality of predictions for different expert co-vary over a pooling space, or as a part of model evaluation.
In this section, we focus on how to create a linear prediction pool \citep*{geweke_optimal_2011} based on these estimates.

\citet{oelrich_modeling_2022} propose using a local linear pool to create aggregate predictions at a new point \(\newobs{\v z}\) in the pooling space
\begin{equation}
    p\left(y \mid \newobs{\v z}\right) = \sum_{k=1}^{K} w_k(\newobs{\v z}) p_k(y), \quad \sum_{k=1}^K w_k(\newobs{\v z}) = 1, \quad w_k(\newobs{\v z}) \ge 0,
\end{equation}
where \(w_k(\newobs{\v z})\), the weight of expert \(k\) at \(\newobs{\mathbf{z}}\), is a function of the posterior probability that expert \(k\) has the highest predictive ability at \(\newobs{\v z}\).
We denote this probability by \(\psi_k(\newobs{\v z}) = p\big(\eta_k(\newobs{\v z}) > \eta_{j \ne k}(\newobs{\v z})\big)\).
A forecast combination using \( \psi_k(\newobs{\v z})\) as weights is referred to as \emph{natural} aggregation in \citet{oelrich_modeling_2022}. 

For aggregation based on single output models --- implicitly assuming independence between experts --- the experimental results in \citet{oelrich_modeling_2022} suggest that the natural weights \(\psi_k(\newobs{\v z})\) do not discriminate strongly enough between experts. \citet{oelrich_modeling_2022} therefore suggest two alternatives.
The first approach, termed \emph{model selection}, gives weight \(1\) to the expert with the highest \(\psi_k(\newobs{\v z})\).
The second approach instead uses a softmax function on the \(\psi_k(\newobs{\v z})\) multiplied by a discrimination factor \(c\):
\begin{equation}\label{softmax_with_discrim}
    w_k(\newobs{\v z}) =
    \frac{
        \exp
        \left( 
            c \cdot \psi_k(\newobs{\v z}) 
        \right)
    }{
        \sum_{j=1}^K 
        \exp
        \left( 
            c \cdot \psi_j(\newobs{\v z})
        \right)
    }.
\end{equation}
This latter approach allows the decision maker to fine-tune how strongly to discriminate between experts based on the posterior probability of having the highest predictive ability.
Setting \(c=\infty\) will lead to the model selection approach, and setting \(c=0\) will lead to equal weights.
Selecting a good value for \(c\) is a problem in itself, and \citet{oelrich_modeling_2022} propose selecting \(c\) dynamically based on past performance for a range of potential \(c\)-values.
This approach is therefore named \emph{dynamic} aggregation.

\begin{figure}
    \centering
    \includegraphics[width=1\textwidth]{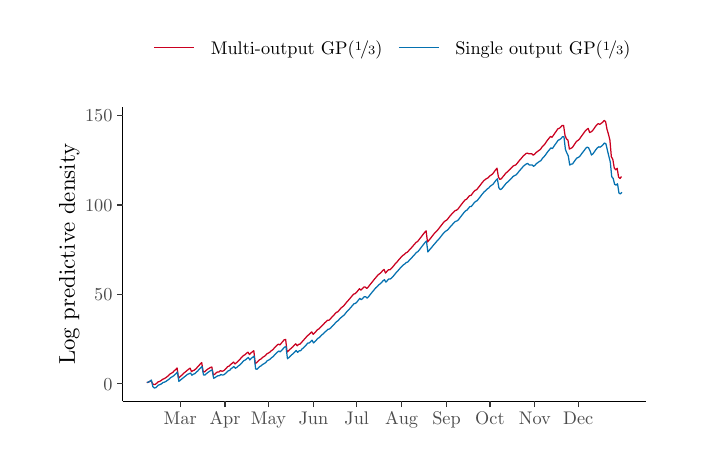}
    \caption{Cumulative log predictive scores based on the natural aggregation scheme for single and multi-output models.}
    \label{fig:aggpreds_thompson_cumu}
\end{figure}

\begin{figure}
    \centering
    \includegraphics[width=1\textwidth]{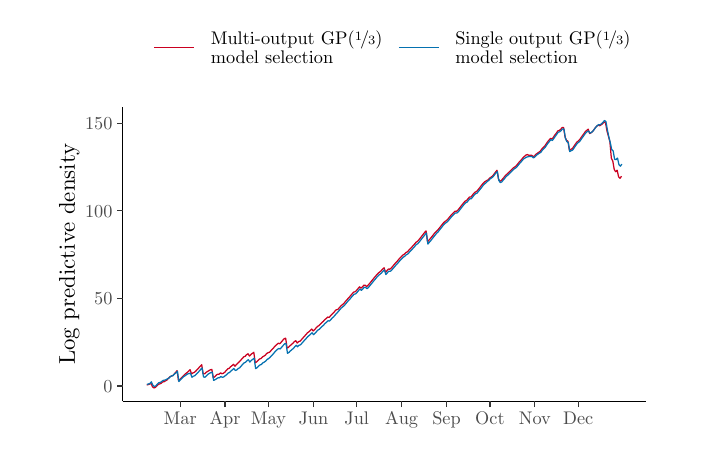}
    \caption{Cumulative log predictive scores based on the model selection aggregation scheme for single and multi-output models.}
    \label{fig:aggpreds_selbest_cumu}
\end{figure}

\begin{figure}
    \centering
    \includegraphics[width=1\textwidth]{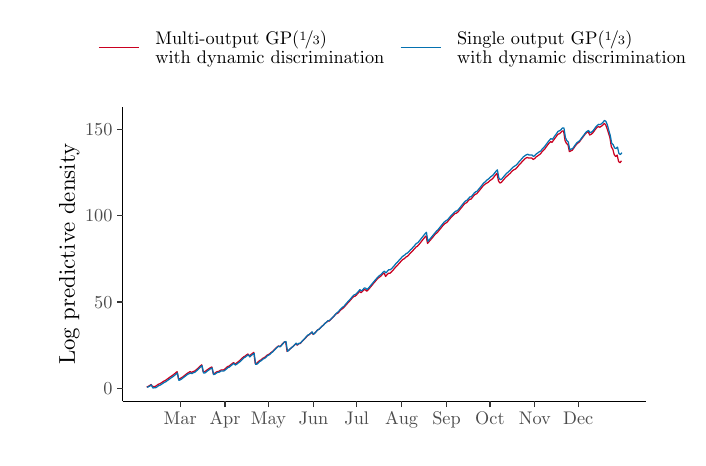}
    \caption{Cumulative log predictive scores based on the dynamic aggregation scheme for single and multi-output models.}
    \label{fig:aggpreds_dynamic_cumu}
\end{figure}

Figures \ref{fig:aggpreds_thompson_cumu}--\ref{fig:aggpreds_dynamic_cumu} compare the three approaches from \citet{oelrich_modeling_2022} with equivalent versions based on the multi-output GP.

The multi-output model clearly outperforms its single output counterparts when using the natural weighting scheme. When using model selection or dynamic discrimination, however, the simpler single-output GP method performs slightly better, in both cases almost entirely due to a few observations at the end of the dataset.

Table \ref{logscore_sums} compares the dynamic multi-output model with a selection of benchmark models from \citet{oelrich_modeling_2022}. Compared to these benchmark models, the multi-output GP's performance puts it between the  global optimal pool \citep{geweke_optimal_2011} and the caliper method \citep{oelrich_local_2021}.

\begin{figure}
    \centering
    \includegraphics[width=12cm]{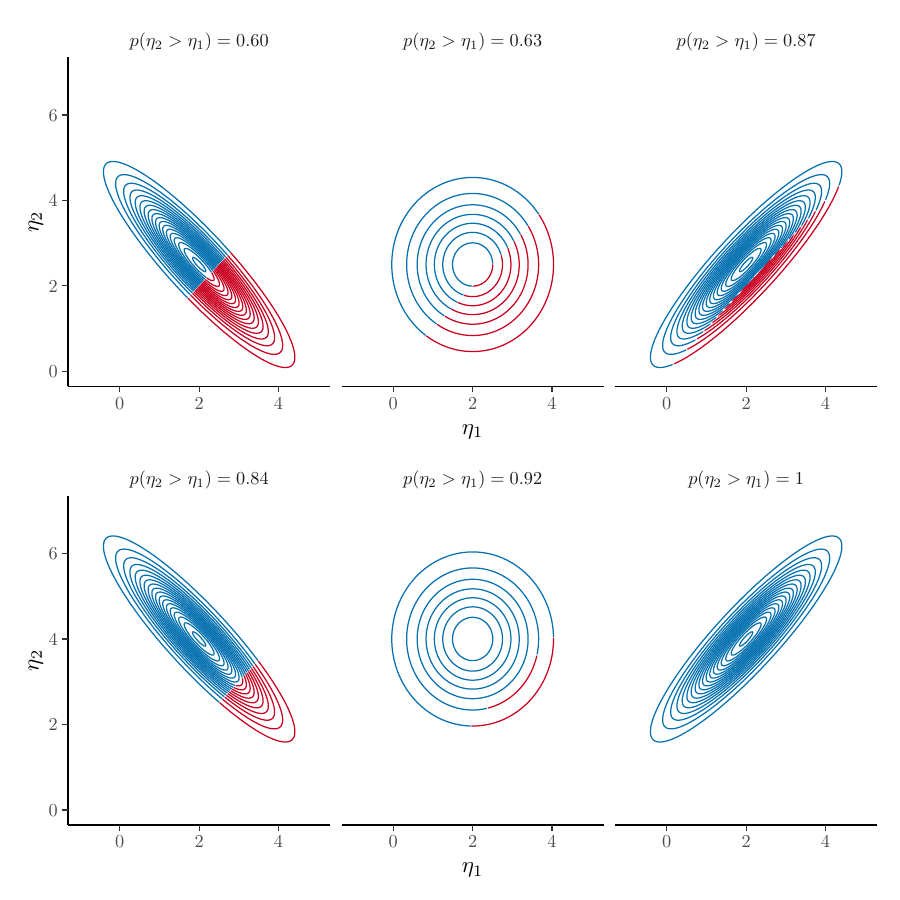}
    \caption{Gaussian contour plot illustrating the effect of correlation in predictive ability on the posterior probability that Expert 2 has greatest predictive ability at $\newobs{\v z}$. In the top row, the marginal difference in posterior mean of predictive ability between the two experts is small ($\E(\eta_2(\newobs{\v z}))-\E (\eta_1(\newobs{\v z})) =0.5$), and in the bottom row the difference is substantial ($\E(\eta_2(\newobs{\v z}))-\E(\eta_1(\newobs{\v z})) =2$). The blue part of each contour plot marks the parts of the joint density where \(\eta_2(\newobs{\v z}) > \eta_{1}(\newobs{\v z})\).}
    \label{fig:correlation_effectc}
\end{figure}

\citet{oelrich_modeling_2022} observe that the degree of discrimination between experts when weighting based on \(\psi_k(\newobs{\v z})\) is often too weak to create good aggregate predictions.
This weak discrimination is in large part caused by a failure to capture correlation between the experts, as a positive correlation will naturally lead to stronger discrimination.
On the other hand, failure to take into account a negative correlation will lead to discrimination that is too strong. This effect is illustrated in Figure \ref{fig:correlation_effectc}.

Figure \ref{fig:correlation_effectc} displays contour plots of the posteriors for $\eta_1$ and $\eta_2$ for two experts in two setups.
In the top row, the difference in posterior mean of predictive ability between the two experts is small (${\E(\eta_2)-\E (\eta_1) =0.5}$), and in the bottom row the difference is more substantial ($\E(\eta_2)-\E(\eta_1) =2$).
In both setups we fix the marginal posterior variances of the two experts at \(1\), and explore what happens when we let the correlation go from strong and negative \((\rho = -0.9)\) to non-existent \((\rho = 0)\) to strong and negative \((\rho = 0.9)\).
As the correlation between the experts approaches one, the posterior probability of having the highest predictive ability will concentrate on the better model, no matter how small the difference in posterior means.
Compared to uncorrelated experts, negative correlation increase the posterior probability that the expert with lower posterior mean has the higher predictive ability, something which becomes clearer as the magnitude of the difference in mean increases.

The effect illustrated in Figure \ref{fig:correlation_effectc} explains why the multi-output GP outperforms the single output version  when using natural weights. By capturing the correlation between experts, the multi-output GP leads to natural weights that discriminate more strongly between the experts. This can also be observed in that the optimal discrimination factor for the multi-output model settles around a lower value than in the single output version. Dynamic discrimination thus allows the single output GP to compensate for the lack of dependence between experts.

\begin{figure}
    \centering
    \includegraphics[scale=0.7]{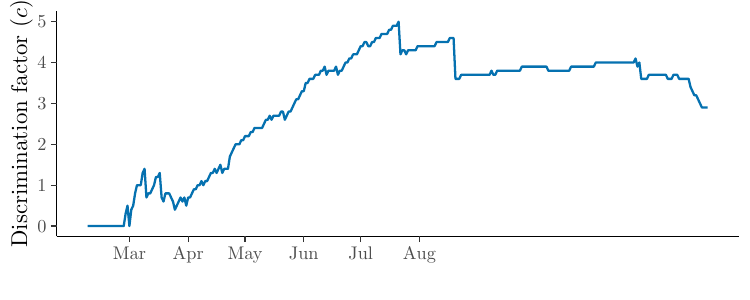}
    \caption{Change in optimal discrimination factor over time.}
    \label{fig:opt_c}
\end{figure}

\begin{table}
\centering
\begin{tabular}{lr}
    \toprule
   Method & Sum of log scores \\ 
   \midrule
   SVBVAR & 45.1 \\ 
   BREG & 28.8 \\ 
   BART & 43.6 \\ 
   Equal weights & 98.7 \\
   Caliper & 133.7 \\ 
   Global opt. & 130.2 \\ 
   Local opt. & 145.3 \\
   \midrule
   $\GP(\nicefrac{1}{3})$ natural & 107.1 \\ 
   multi-$\GP(\nicefrac{1}{3})$ natural & 115.9 \\ 
   $\GP(\nicefrac{1}{3})$ model sel. & 126.6 \\ 
   multi-$\GP(\nicefrac{1}{3})$ model sel. & 119.7 \\
   $\GP(\nicefrac{1}{3})$ dynamic & 136.4 \\
   multi-$\GP(\nicefrac{1}{3})$ dynamic & 131.7 \\
     \bottomrule
  \end{tabular}
  \caption{Sum of log predictive density scores over the whole test period.}
  \label{logscore_sums}
\end{table}

\section{Conclusions and further research}\label{sec_conclusions}

We extend the method for estimating the local predictive ability of an expert of \citet{oelrich_modeling_2022} to the joint estimation of the local predictive ability of a set of experts, using a multi-output Gaussian process.
The proposed model makes it possible to capture correlation between the predictive abilities of the experts as well as correlation in the noise.

We apply the power transformation suggested in \citet{oelrich_modeling_2022} to the log scores of each expert and show that this allows us to integrate out the multivariate GP surface analytically to obtain the marginal posterior of the hyperparameters. This marginalization approach makes inference computationally tractable for the joint posterior of the local predictive abilities of all experts. We demonstrate the advantages of this approach in simulations, and apply it to the prediction of bike sharing data where it is able to capture correlations between experts, but performs similarly to using multiple single output GPs.

While the inference and prediction from the marginalized version is much faster compared to sampling also the latent variables in the Multi-GP($\nicefrac{1}{3}$) model, the step from running several single output GPs side by side to running a single multi-output GP increases the computational burden significantly, making large sample sizes time consuming.
A topic for further study is methods for speeding up the calculation using sparse linear algebra, methods for large-scale homoscedastic Gaussian GPs \citep{liu2020gaussian}, or using alternative Bayesian inference approaches than Monte Carlo sampling, such as variational inference based on optimization \citep{blei2017variational}.

Another potential avenue of future research is using alternative methods to pool predictions based on local predictive ability. In the empirical application we use three different methods, all of which have pooling weights based on \(\psi_k\), the posterior probability that a particular expert has greater predictive ability than all other experts. Alternatively, one can form linear combinations directly on the $\eta_1(\v z),\ldots,\eta_K(\v z)$ to obtain a posterior for the pooled local ability $\sum_{k=1}^K w_k \eta_k(\v z)$. 
Given a set of criteria for this distribution, such as maximizing the mean while keeping the variance below a certain threshold, a set of local weights can be obtained through numerical optimization.

\section*{}

\bibliographystyle{apalike}
\bibliography{my_lib}

\end{document}